\documentclass{article}
\usepackage[dvips]{graphicx}
\usepackage{spconf,amssymb,amsmath,epsfig,color}
\newcommand{\beq}{\begin{equation}}
\newcommand{\eeq}{\end{equation}}
\newcommand{\beqn}{\begin{eqnarray}}
\newcommand{\eeqn}{\end{eqnarray}}
\DeclareMathOperator*{\argmin}{arg\,min}
\def\bmath#1{\mbox{\boldmath$#1$}}

\long\def\symbolfootnote[#1]#2{\begingroup%
\def\thefootnote{\fnsymbol{footnote}}\footnote[#1]{#2}\endgroup}

\title{Fine tuning consensus optimization for distributed radio interferometric calibration}
\name{Sarod Yatawatta\thanks{This work is supported by the European Research Council project LOFARCORE, grant \#339743.}}
\address{ASTRON, The Netherlands Institute for Radio Astronomy,\\ The Netherlands.\\ Email: yatawatta@astron.nl}

\begin{document}
%
\maketitle
\begin{abstract}
We recently proposed the use of consensus optimization as a viable and effective way to improve the quality of calibration of radio interferometric data. We showed that it is possible to obtain far more accurate calibration solutions and also to distribute the compute load across a network of computers by using this technique. A crucial aspect in any consensus optimization problem is the selection of the penalty parameter used in the alternating direction method of multipliers (ADMM) iterations. This affects the convergence speed as well as the accuracy. In this paper, we use the Hessian of the cost function used in calibration to appropriately select this penalty. We extend our results to a  multi-directional calibration setting, where we propose to use a penalty scaled by the squared intensity of each direction.
\end{abstract}
\section{Introduction}
Modern radio interferometric arrays deliver large volumes of data, in order to reach higher sensitivities yielding new science. To reach the full potential of such arrays, estimation of systematic errors in the data and correction for such errors (also called as calibration) is essential. This is not a trivial task for an array with hundreds of receivers that collect data over many hours and at thousands of different frequencies. A case in point being the square kilometre array (SKA), which is in the planning phase. Thus, there is an urgent need for computationally efficient and robust algorithms.  On the other hand, there is a surge in research related to large scale and distributed data processing algorithms (also called as big-data), which we can exploit to solve some of these problems. 

Our recent work \cite{DCAL} introduced distributed-calibration as a way of distributing the computational burden over a network of computers while at the same time improving the quality of calibration. We essentially exploited the continuity of systematic errors over frequency to enforce an additional constraint onto calibration. This reduces calibration to a consensus optimization \cite{boyd2011} problem and we used alternating direction method of multipliers (ADMM) \cite{BT}  as the underlying algorithm in the proposed distributed calibration scheme.

 Consensus optimization, practically implemented with ADMM, has been extensively studied and is deployed in a wide variety of application areas (some recent examples are \cite{Chang2014,Wei2012,Erseghe12}). In addition, similar work is beginning to appear in radio astronomical imaging \cite{Ferrari2014,PURIFY,Onose}.  However, compared with other users of ADMM, we observe several unique properties of the calibration problem that we face. First, the cost function used in calibration is non-linear and non-convex.  The systematic errors are mainly caused by directional effects such as the ionosphere and the receiver beam shape. Although we know the general properties of such errors, building an entirely accurate model (for instance for their variation with frequency) is not feasible. Hence, we enforce consensus only by using an approximate model, and this is clearly different and also more involved from most other applications. Indeed, other applications such as consensus averaging, where consensus is enforced on a constant value, use a perfect model. Furthermore, most other applications use complicated network topologies (that in turn affect the performance of ADMM) and on the other hand, in our case, we have a much simpler (and fully connected) network with one fusion center.

Of particular interest is the convergence rate of ADMM, which depends on many factors including the penalty parameter and the network topology \cite{nishihara2015general}. In most cases, the penalty parameter is selected by trial and error, following some general guidelines \cite{BT}. However, for specific problems, better methods to select the penalty have been proposed \cite{nishihara2015general,Teix2016,Ghadimi2015}. Recent work \cite{Hong15} has suggested to select the penalty parameter as large as possible to make the objective function strongly convex. Hence for our problem, we study the Hessian of the cost function to select appropriate values for the penalty parameter. For calibration along multiple directions in the sky, we can select different penalty values along each direction. Intuitively, we select a large penalty along directions with higher signal where we have more confidence in our model. These directions are mostly close to the center of the field of view. On the other hand, for directions far away from the center, we select a smaller penalty.  

The rest of the paper is organized as follows: In section \ref{sec:calib} we give an overview of radio interferometric calibration. Next, in section \ref{sec:dist}, we present distributed calibration based on consensus optimization. We also present a scheme based on the Hessian of the cost function to select the penalty parameter. Simulation results are presented in section \ref{sec:results} where we demonstrate the improved performance with a refined  penalty parameter. Finally, we draw our conclusions in section \ref{sec:conclusions}.

Notation: Matrices and vectors are denoted by bold upper and lower case letters as ${\bf J}$ and ${\bf v}$, respectively. The transpose and the  Hermitian transpose are given by $(.)^T$ and $(.)^H$. The matrix  Frobenius norm is given by $\|.\|$. The set of real and complex numbers are denoted by  ${\mathbb R}$ and ${\mathbb C}$. The identity matrix is given by $\bf I$. The matrix trace operator is given by $\rm{trace}(.)$. 
\section{Radio Interferometric Calibration}\label{sec:calib}
Consider a radio interferometric array with $N$ receivers. The sky is composed of many discrete sources and we consider calibration along $K$ directions in the sky. The observed data at a baseline formed by two receivers, $p$ and $q$ is given by
\cite{HBS}
\beq \label{ME}
{\bf V}_{pq}=\sum_{k=1}^K{\bf J}_{pk} {\bf C}_{pqk} {\bf J}_{qk}^H + {\bf N}_{pq}
\eeq
where ${\bf V}_{pq}$ ($\in \mathbb{C}^{2\times 2}$) is the observed {\em visibility} matrix (or the cross correlations). The systematic errors that need to be calibrated for station $p$ and $q$ are given by the Jones matrices ${\bf J}_{pk},{\bf J}_{qk}$ ($\in \mathbb{C}^{2\times 2}$), respectively. Note that since $K$ directions are calibrated, for each station, there are $K$ Jones matrices (so $KN$ in total). The sky signal (or {\em coherency}) along the $k$-th  direction is given by ${\bf C}_{pqk}$ ($\in \mathbb{C}^{2\times 2}$) and is known a priori. The values of ${\bf J}_{pk},{\bf J}_{qk}$ and ${\bf C}_{pqk}$ in (\ref{ME}) are implicitly dependent on sampling time and frequency of the observation. The noise matrix ${\bf N}_{pq}$ ($\in \mathbb{C}^{2\times 2}$) is assumed to have complex, zero mean, circular Gaussian elements.

Estimating the Jones matrices in (\ref{ME}) can be further simplified by using the space alternating generalized expectation maximization (SAGE) algorithm \cite{Fess94,Kaz2}. In a nutshell, using SAGE algorithm, we can simplify calibration along $K$ directions to $K$ single direction calibration subproblems (see \cite{Kaz2} for details). Calibration along the $k$-th direction is done by using the effective observed data
\beq \label{ME1}
{\bf V}_{pqk} = {\bf V}_{pq} - \sum_{l=1,l\ne k}^K\widehat{\bf J}_{pl} {\bf C}_{pql} \widehat{\bf J}_{ql}^H
\eeq 
using current estimates $\widehat{\bf J}_{pl}$ and $\widehat{\bf J}_{ql}$ and for an array with $N$ receivers, we can form at most $N(N-1)/2$ baselines that collect visibilities as in (\ref{ME1}), for any given time and frequency sample. We define our objective function (for the $k$-th direction) under a Gaussian noise model as
\beq \label{cost1}
g_{k}({\bf J}_{1k},{\bf J}_{2k},\ldots)= \sum_{p,q}\| {\bf V}_{pqk} - {\bf J}_{pk} {\bf C}_{pqk} {\bf J}_{qk}^H \|^2
\eeq
where the summation is over the baselines $pq$ that have data. By increasing the time and frequency interval within which data are collected, this summation can be expanded (thus improving the signal to noise ratio).
By defining ${\bf J}$ ($\in \mathbb{C}^{2N\times 2}$) as  the augmented matrix of Jones matrices of all stations along the $k$-th direction,
\beq
{\bf J}\buildrel\triangle\over=[{\bf J}_{1k}^T,{\bf J}_{2k}^T,\ldots,{\bf J}_{Nk}^T]^T,
\eeq
and ${\bf A}_p$ ($\in \mathbb{R}^{2\times 2N}$) (and ${\bf A}_q$ likewise) as the canonical selection matrix
\beq \label{Ap}
{\bf A}_p \buildrel\triangle\over=[{\bf 0},{\bf 0},\ldots,{\bf I},\ldots,{\bf 0}],
\eeq
(only the $p$-th block of (\ref{Ap}) is an identity matrix) we can rewrite (\ref{cost1}) as
\beq \label{cost2}
g_{k}({\bf J})= \sum_{p,q}\| {\bf V}_{pqk} - {\bf A}_p{\bf J} {\bf C}_{pqk} ({\bf A}_q{\bf J})^H \|^2.
\eeq

Calibration along the $k$-th direction is the estimation of ${\bf J}$ by minimizing  (\ref{cost2}). Note that (\ref{cost2}) has to be minimized for each direction $k=1\ldots K$ and updated values of (\ref{ME1}) are re-used until convergence is reached in the SAGE algorithm. We also note that (\ref{cost2}) only gives solutions for one frequency and time interval, and to calibrate the full dataset, many such solutions are obtained for data observed at different time and frequency intervals.

\section{Distributed Calibration}\label{sec:dist}
We have introduced calibration along $K$ directions, but only working on a single frequency and time sample in section \ref{sec:calib}. In this section, we consider calibrating data observed at $P$ different frequencies, but only along $1$ direction, because this can easily be extended to $K$ directions using the SAGE algorithm. We impose an additional constraint that tries to preserve continuity of ${\bf J}$ in (\ref{cost2}) over frequency. To solve this, we introduced the use of consensus optimization in \cite{DCAL}, where the objective function is modified into an augmented Lagrangian  
\beq \label{aug}
L_f({\bf J}_f,{\bf Z},{\bf Y}_f)=g_{f}({\bf J}_f) + \|{\bf Y}_f^H({\bf J}_f-{\bf B}_f {\bf Z})\| + \frac{\rho}{2} \|{\bf J}_f-{\bf B}_f {\bf Z}\|^2
\eeq
where the subscript $(.)_f$ denotes data (and parameters) at frequency $f$. In (\ref{aug}), $g_{f}({\bf J}_f)$ is the original cost function as in (\ref{cost2}), except that the subscripts denote frequency $f$. The Lagrange multiplier is given by ${\bf Y}_f$ ($\in \mathbb{C}^{2N\times 2}$). The calibration parameters are given by ${\bf J}_f$ ($\in \mathbb{C}^{2N\times 2}$). The continuity in frequency is enforced by the frequency model given by ${\bf B}_f$ ($\in \mathbb{R}^{2N\times 2NF}$), which is essentially a set of basis functions in frequency, evaluated at $f$. The global variable ${\bf Z}$ ($\in \mathbb{C}^{2NF\times 2}$) is shared by data at all $P$ frequencies.

The ADMM iterations for solving (\ref{aug}) are given as
\beqn \label{step1}
({\bf J}_f)^{n+1}= \underset{{\bf J}}{\argmin}\ \ L_f({\bf J},({\bf Z})^n,({\bf Y}_f)^n)\\ \label{step2}
({\bf Z})^{n+1}= \underset{{\bf Z}}{\argmin}\ \ \sum_f L_f(({\bf J}_f)^{n+1},{\bf Z},({\bf Y}_f)^n)\\ \label{step3}
({\bf Y}_f)^{n+1}=({\bf Y}_f)^n + \rho\left( ({\bf J}_f)^{n+1}-{\bf B}_f ({\bf Z})^{n+1} \right)
\eeqn
where we use the superscript $(.)^n$ to denote the $n$-th iteration. The steps (\ref{step1}) and (\ref{step3}) are done for each $f$ in parallel. The update of the global variable (\ref{step2}) is done at the fusion center. More details of these steps can be found in \cite{DCAL}.  

In this paper, we study strategies for selecting the penalty parameter $\rho$ to get faster convergence and accurate results. In order to do this, we use the Hessian operator of the cost function (\ref{cost2}), which is given as \cite{DCAL,ICASSP13},
\beqn\label{Hess}
\lefteqn{\mathrm{Hess}_f\left(g_{f}({\bf {J}}),{\bf {J}},{\bmath \eta}\right)}\\\nonumber
&=&\sum_{p,q}\left( {\bf {A}}_p^T \left( ({\bf {V}}_{pqf}-{\bf {A}}_p{\bf {J}}{\bf {C}}_{pqf}{\bf {J}}^H{\bf {A}}_q^T) {\bf {A}}_q {\bmath \eta}\right.\right.\\\nonumber
&& \left.\left.- {\bf {A}}_p({\bf {J}}{\bf {C}}_{pqf} {\bmath \eta}^H + {\bmath \eta}{\bf {C}}_{pqf}{\bf {J}}^H) {\bf {A}}_q^T{\bf {A}}_q{\bf {J}}\right) {\bf {C}}_{pqf}^H\right. \\\nonumber
&&\left. + {\bf {A}}_q^T \left( ({\bf {V}}_{pqf}-{\bf {A}}_p{\bf {J}}{\bf {C}}_{pqf}{\bf {J}}^H{\bf {A}}_q^T)^H {\bf {A}}_p {\bmath \eta}\right.\right.\\\nonumber
&& \left.\left.- {\bf {A}}_q({\bf {J}}{\bf {C}}_{pqf} {\bmath \eta}^H + {\bmath \eta}{\bf {C}}_{pqf}{\bf {J}}^H)^H {\bf {A}}_p^T{\bf {A}}_p{\bf {J}}\right) {\bf {C}}_{pqf}\right) \\\nonumber
\eeqn
where ${\bmath \eta}\in \mathbb{C}^{2N\times 2}$.

For convexity, we need a positive definite Hessian. Since we have a Hessian operator (instead of a matrix), we need to find the smallest eigenvalue of the Hessian, and for convexity, this should be positive. In order to find this, we define a cost function as
\beqn \label{hcost}
\lefteqn{h({\bmath \eta})\buildrel\triangle\over= \frac{1}{2}\mathrm{trace}\left({\bmath \eta}^H \mathrm{Hess}_f\left(g_{f}({\bf {J}}),{\bf {J}},{\bmath \eta}\right)\right.}\\\nonumber
&&+\left.\mathrm{Hess}_f^H\left(g_{f}({\bf {J}}),{\bf {J}},{\bmath \eta}\right) {\bmath \eta}\right)
\eeqn
and we  find the smallest eigenvalue $\lambda$ by solving
\beqn \label{eig}
&&\lambda=\underset{{\bmath \eta}}{\argmin}\ \ \ \ h({\bmath \eta})\\\nonumber
&&{\mathrm{subject\ to}}\ \ {\bmath \eta}^H{\bmath \eta}={\bf I}.
\eeqn

The constraint ${\bmath \eta}^H {\bmath \eta}={\bf I}$ makes the minimization of (\ref{hcost}) restricted onto a complex Stiefel manifold \cite{AMS}, which can be easily solved by using the Riemannian trust region method \cite{RTR,manopt}. In order to do this, we require the gradient and Hessian of $h({\bmath \eta})$, which are given as
\beq
\mathrm{grad}\left(h({\bmath \eta}),{\bmath \eta}\right)= \mathrm{Hess}_f\left(g_{f}({\bf {J}}),{\bf {J}},{\bmath \eta}\right)
\eeq
 and 
\beq
\mathrm{Hess}\left(h({\bmath \eta}),{\bmath \eta},{\bmath \zeta}\right)=  \mathrm{Hess}_f\left(g_{f}({\bf {J}}),{\bf {J}},{\bmath \zeta}\right),
\eeq
where ${\bmath \zeta} \in \mathbb{C}^{2N\times 2}$.

After obtaining $\lambda$ from (\ref{eig}), our strategy is to select $\rho$ such that $\rho+\lambda\ge 0$ so that the Hessian of the augmented Lagrangian (\ref{aug}) is positive semi-definite \cite{Hong15}. In order to do this, we need an estimate for ${\bf J}$ in (\ref{hcost}). We can find this by initial calibration with a pre-determined value of $\rho$ (say $\rho=0$). Once we obtain $\widehat{\bf J}$,  we use  (\ref{eig}) to find $\lambda$ and afterwards we update $\rho$. Note that $\lambda$ is dependent on $f$, but we ignore the frequency dependence of $\lambda$ and use one value of $f$ (typically the middle) to estimate it. 

So far, we have considered calibration along one direction only. The next question that we must answer is how to select $\rho$ for calibration along $K$ directions in the sky. For each direction, ${\bf {C}}_{pqf}$ in (\ref{Hess}) will influence the value of $\lambda$. If the centroid of the source (cluster) \cite{Kazemi3}  is along $l,m$ direction in the sky and if its effective (unpolarized) intensity is $\alpha$, we have
\beq \label{coh}
{\bf {C}}_{pqf} \approx \exp\left(\jmath \phi(l,m,p,q)\right) \alpha {\bf I}
\eeq
where $\phi(l,m,p,q)$ is the phase contribution and ${\bf I}$ is a $2\times 2$ identity matrix. Hence ${\bf {C}}_{pqf}$ is a diagonal scalar matrix. If $\widehat{\bf J}$ is close to the true solution, the term ${\bf {V}}_{pqf}-{\bf {A}}_p{\bf {J}}{\bf {C}}_{pqf}{\bf {J}}^H{\bf {A}}_q^T$ becomes negligible compared with the other terms in (\ref{Hess}). The remaining terms have a product ${\bf {C}}_{pqf} {\bf {C}}_{pqf}^H$ and the phase term in (\ref{coh}) cancel out. Therefore, for different clusters, the value for $\lambda$ obtained by (\ref{eig}) is mainly determined by the squared effective intensity $\alpha^2$ of each source. Hence, once we have determined a suitable value for $\rho$ for one direction, the corresponding values for other directions can be determined by scaling by the squared effective intensity.
\section{Simulation Results}\label{sec:results}
We simulate an array of $N=47$ receivers that calibrate along $K=5$ directions in the sky. The matrices ${\bf J}_{pk},{\bf J}_{qk}$ in (\ref{ME}) are generated with their elements having values drawn from a complex uniform distribution in $[0,1]$, multiplied by a frequency dependence given by a random $7$-th order polynomial. The intensities of the $K=5$ sources are randomly generated in the range $[1,5]$ and their positions are randomly chosen in a field of view of about $7\times 7$ square degrees. The variation of intensities with frequency is given by a power law with randomly generated exponent in $[-1,1]$. The noise matrices ${\bf N}_{pq}$ in (\ref{ME}) are simulated to have complex circular Gaussian random variables. The variance of the noise is changed according to the signal to noise ratio ($\rm{SNR}=10$) 
\beq
\mathrm{SNR}\buildrel\triangle\over=\frac{\sum_{p,q} \|{\bf V}_{pq}\|^2}{\sum_{p,q} \| {\bf N}_{pq}\|^2}.
\eeq
With this setup, we generate data for $P=8$ frequency channels in the range $115$ to $185$ MHz. For calibration, we setup a $3$-rd order polynomial model ($F=4$), using Bernstein basis functions \cite{Farouki} for the matrix ${\bf B}_f$ in (\ref{aug}). Note that we intentionally use a lower order frequency dependence than what is actually present in the data to create a realistic scenario when the exact model is not known. During calibration, initial values for the parameters are always set as ${\bf J}_p={\bf I}$ for $p\in[1,N]$. Unless stated otherwise, all directions have the same value of $\rho$. We use $50$ ADMM iterations, and after the $1$-st iteration, we solve (\ref{eig}) to estimate $\lambda$, and we get a typical value of $\lambda=-150$ for a source with unit amplitude. Regardless, we perform calibration with various values of $\rho$ to compare performance.

We find the normalized (averaged over all directions) mean squared error (NMSE) between true ${\bf J}_f$ and its estimate as
\beq\label{nmse}
\mathrm{NMSE}\buildrel\triangle\over=\frac{1}{\sqrt{2KN}}\sqrt{\sum_k \|{\bf {J}}_f-\widehat{\bf {J}}_f {\bf {U}}\|^2}
\eeq
to measure the accuracy of calibration. In (\ref{nmse}), ${\bf U}$ is a unitary matrix that removes the unitary ambiguity in the estimated $\widehat{\bf J}_f$ \cite{interpolation}.

\begin{figure}[htbp]
\begin{minipage}[b]{0.98\linewidth}
\begin{minipage}[b]{0.48\linewidth}
\centering \centerline{\epsfig{figure=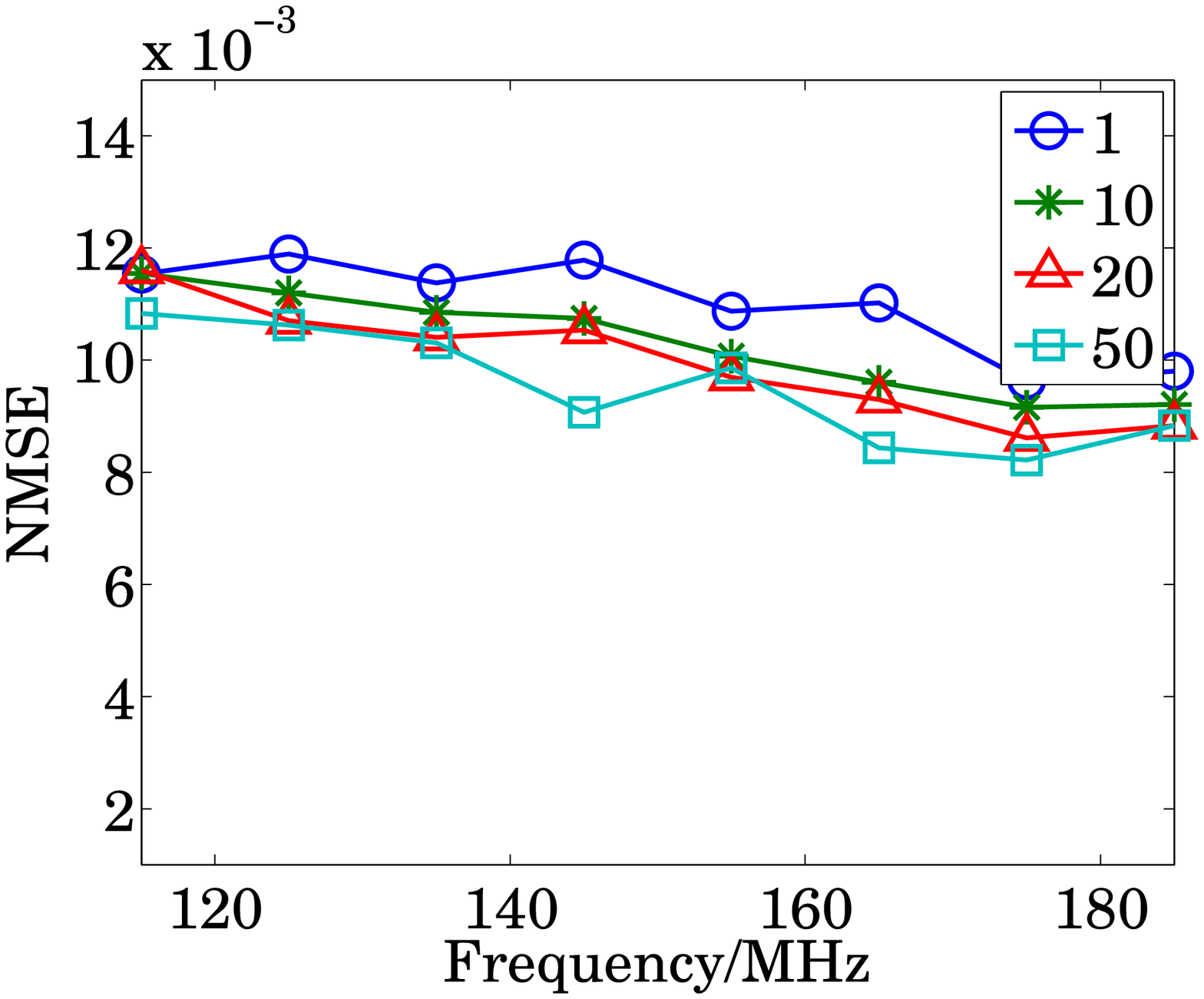,width=4.2cm}}
\vspace{0.2cm}\centerline{$\  \rho=5$}
\end{minipage}
\begin{minipage}[b]{0.48\linewidth}
\centering \centerline{\epsfig{figure=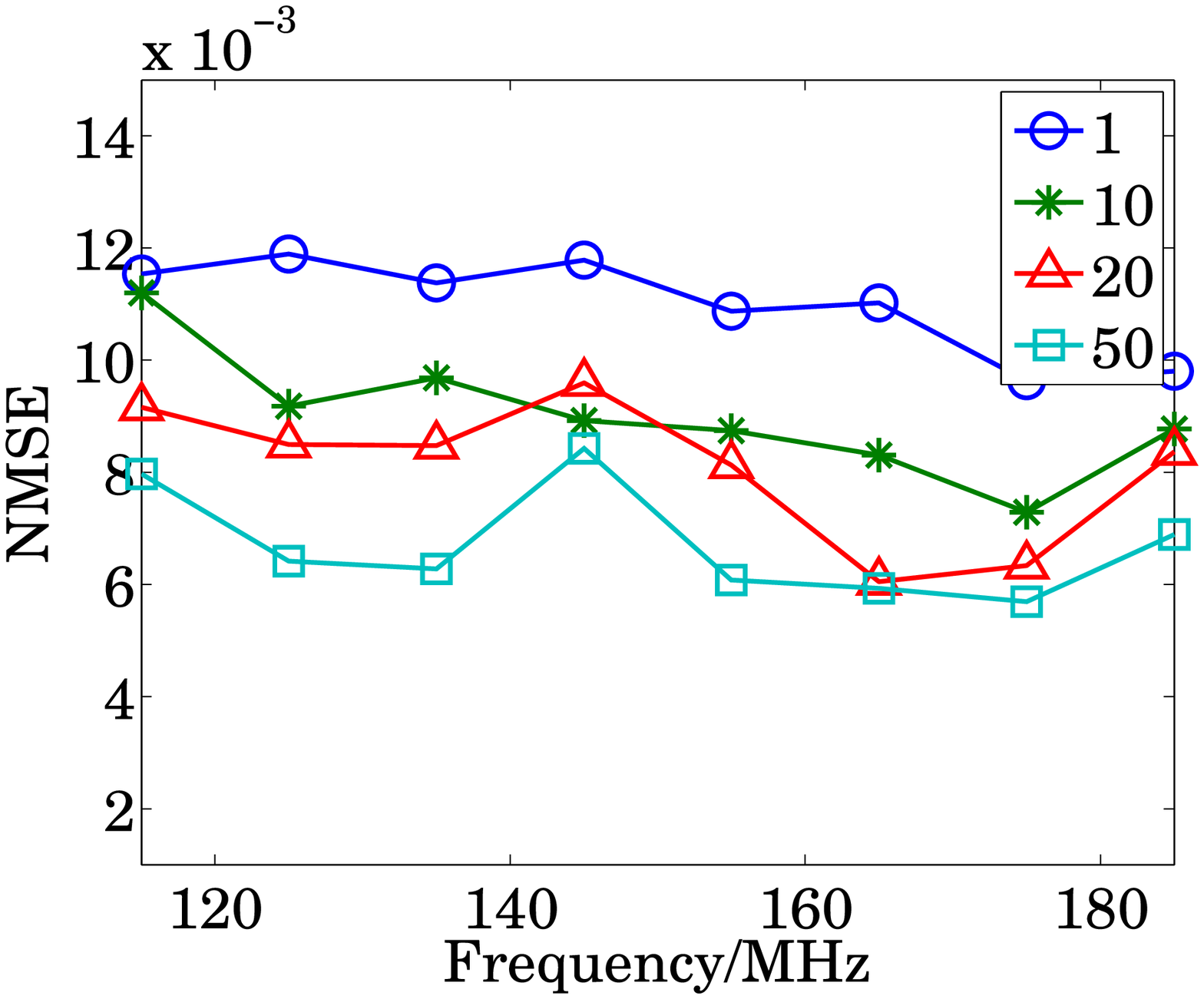,width=4.2cm}}
\vspace{0.2cm}\centerline{$\rho=50$}
\end{minipage}
\begin{minipage}[b]{0.48\linewidth}
\centering \centerline{\epsfig{figure=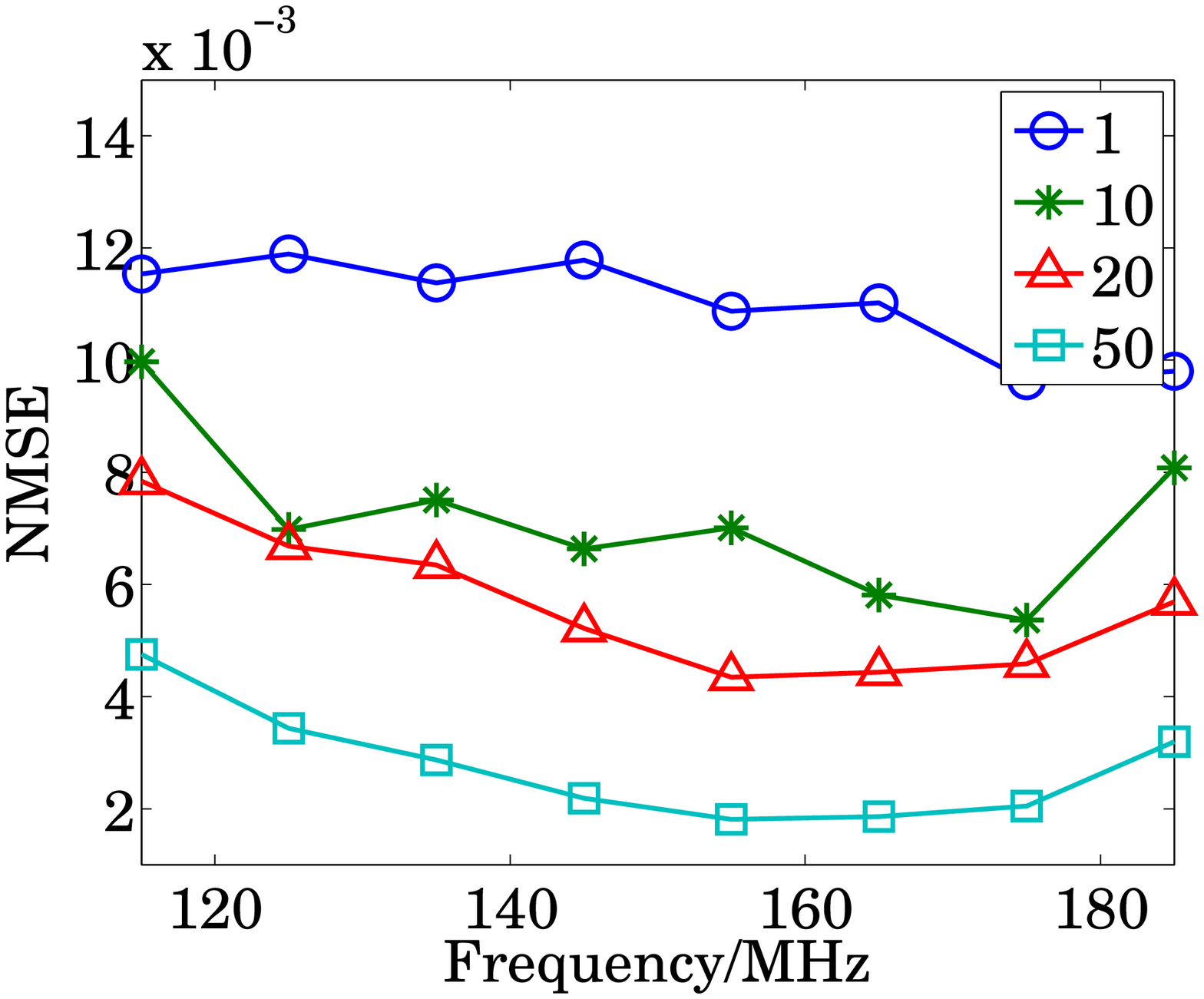,width=4.2cm}}
\vspace{0.2cm}\centerline{$\rho=200$}
\end{minipage}
\begin{minipage}[b]{0.48\linewidth}
\centering \centerline{\epsfig{figure=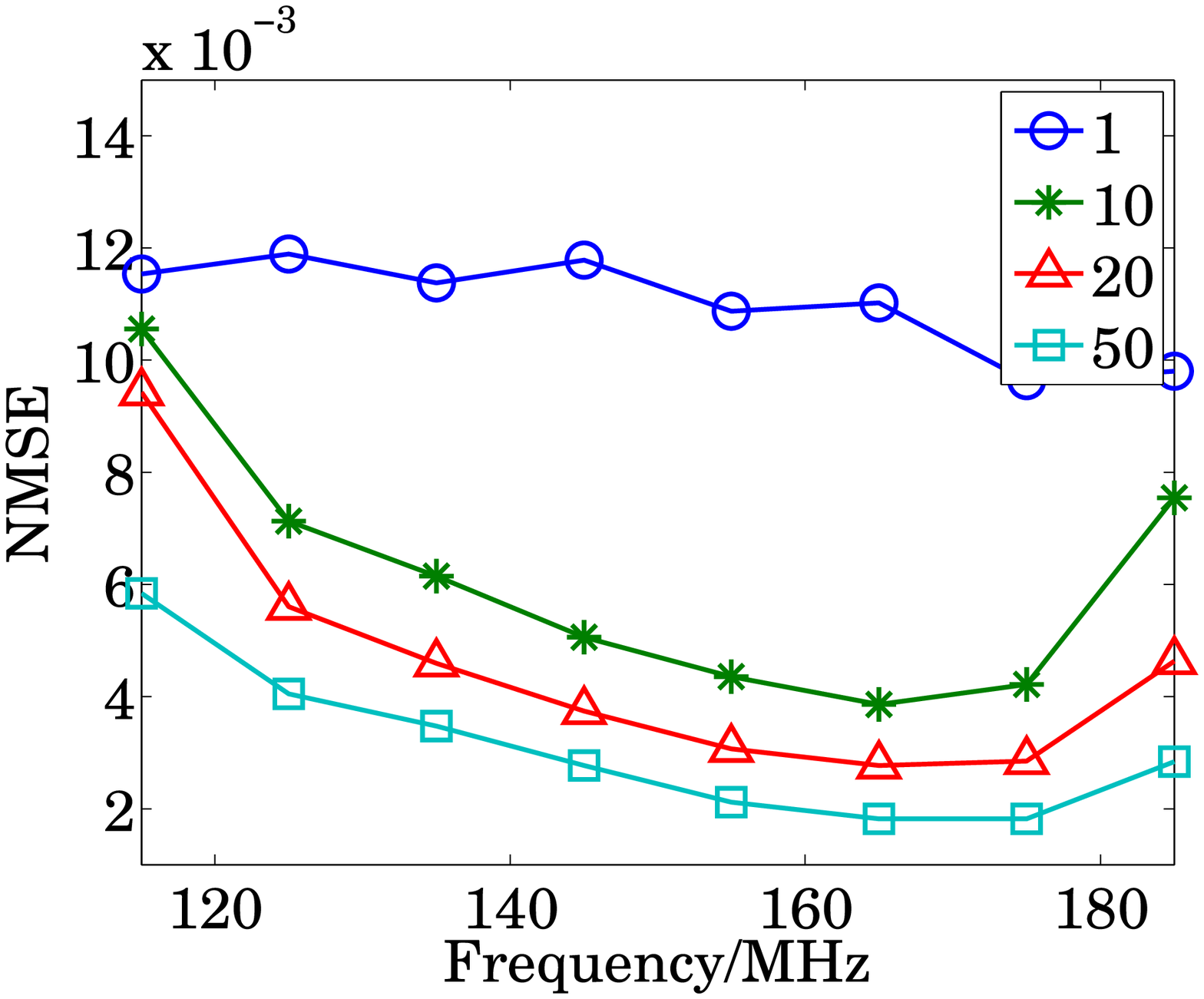,width=4.2cm}}
\vspace{0.2cm}\centerline{$\rho=1000$}
\end{minipage}
\end{minipage}
\caption{NMSE for various $\rho$ with increasing ADMM iterations.}
\label{fignmse}
\end{figure}

\begin{figure}[htbp]
\begin{minipage}[b]{0.98\linewidth}
\centering
\centerline{\epsfig{figure=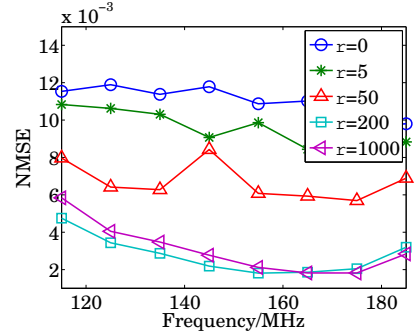,width=5.4cm}}
\end{minipage}
\caption{NMSE for various $\rho$ after $50$ ADMM iterations.} \label{fignmseall}
\end{figure}

In Fig. \ref{fignmse}, we show the NMSE for various values of $\rho$, with increasing number of ADMM iterations. We see that for $\rho+\lambda>0$ ($\rho=200$) we get the best performance, but increasing $\rho$ too much beyond this value ($\rho=1000$) shows no additional improvement. A notable behavior of the NMSE is the enhancement of the error at the edges (especially at low ADMM iterations), which we attribute to Runge's phenomenon \cite{Runge} in polynomial interpolation.
In Fig. \ref{fignmseall}, we show the final NMSE for 50 ADMM iterations, which once again shows that $\rho=200$ gives the best result.
\begin{figure}[htbp]
\begin{minipage}[b]{0.98\linewidth}
\centering
\centerline{\epsfig{figure=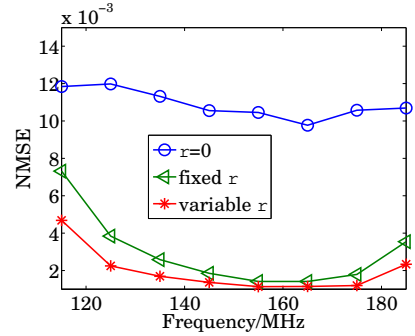,width=5.4cm}}
\end{minipage}
\caption{NMSE after $50$ ADMM iterations with fixed $\rho$ along all directions and varying $\rho$ according to squared intensity.} \label{fignmsevar}
\end{figure}

In Fig. \ref{fignmsevar}, we show NMSE for a simulation with intensities at mid frequency $5,3,3,2$ and $1.5$ along the $K=5$ directions. In one calibration, we use regularization $\rho=400$ for all directions and in the other, we use $\rho$ equal to $400,144,144,64$ and $36$ respectively. We see that varying $\rho$ in proportion to the squared intensity gives the better NMSE.

\section{Conclusions}\label{sec:conclusions}
We have investigated refining the performance of distributed calibration based on consensus optimization in this paper. We have used the Hessian of the cost function to appropriately select the penalty parameter such that the augmented Lagrangian becomes convex. Furthermore, in a multi-directional calibration scheme, we have proposed to scale the penalty parameter proportional to the squared intensity along each direction. According to our simulations, such fine-tuning of parameters gives superior performance in terms of accuracy and convergence of the distributed calibration scheme.
\bibliographystyle{IEEE}
\bibliography{references}

\end{document}